\documentclass{article}
\usepackage{spconf,amsmath,graphicx,amsfonts,amssymb}
\usepackage[export]{adjustbox}

\title{Improving Music Genre Classification from multi-modal properties of music and genre correlations Perspective}

\name{Ganghui Ru$^\ddag$, Xulong Zhang$^\ddag$, Jianzong Wang$^\ast$, Ning Cheng, Jing Xiao\thanks{$\ddag$These authors contributed equally to this work.} 
\thanks{$^\ast$Corresponding author: Jianzong Wang, jzwang@188.com.}}
\address{Ping An Technology (Shenzhen) Co., Ltd., China}
\begin{document}
%\ninept
%

\maketitle

\begin{abstract}
Music genre classification has been widely studied in past few years for its various applications in music information retrieval. Previous works tend to perform unsatisfactorily, since those methods only use audio content or jointly use audio content and lyrics content inefficiently. In addition, as genres normally co-occur in a music track, it is desirable to capture and model the genre correlations to improve the performance of multi-label music genre classification. To solve these issues, we present a novel multi-modal method leveraging audio-lyrics contrastive loss and two symmetric cross-modal attention, to align and fuse features from audio and lyrics. Furthermore, based on the nature of the multi-label classification, a genre correlations extraction module is presented to capture and model potential genre correlations. Extensive experiments demonstrate that our proposed method significantly surpasses other multi-label music genre classification methods and achieves state-of-the-art result on Music4All dataset.
\end{abstract}
\section{Introduction}
\label{sec:intro}
With the growing prosperity of digital music market, how to correctly classify music genres has become an urgent problem. In various tasks of music information retrieval (MIR), music genre classification (MGC) is one of the oldest and most important tasks, and has become a research hotspot due to its wide application prospects. However, since the complex nature and diverse expressions of music tracks, music usually has more than one genre category, characterizing a multi-label classification problem. Compared with conventional MGC task, multi-label MGC is more challenging due to the combinatorial nature of the output space. 

Early studies \cite{early_work1,early_work3} on MGC task were based on digital signal processing and traditional machine learning methods, which use some hand-crafted features as inputs. Limited by the ability of representation, it is difficult for these methods to extract deep and abstract information. Recently, methods based on deep learning gradually become mainstream, showing great potential. Some studies \cite{lyrics,cnn,rnn} contributed to MGC task using CNN or RNN. These methods usually used single information (uni-modal) such as audio or lyrics, thus leading to prone to performance bottlenecks. With the rapid development of computing resources, many MIR tasks have evolved from uni-modal to multi-modal. Some studies \cite{multi-modal1,multi-modal2, multi-modal3} utilized both audio content and text content of a music track for music genre classification. Especially, \cite{zhao} proposed a multi-modal method with the hierarchical cross-attention network for music emotion recognition. In \cite{six_feature}, Vatolkin et al. analyzed the impact of six different modalities on the MGC task, including audio signals, lyric texts, album cover images, audio semantic tags, playlist co-occurrences, and symbolic MIDI representations. Although the methods mentioned above have achieved certain success by simple fusion of multi-modal features, they ignore that features from different modalities are inherently heterogeneous and misaligned. To address this issue, we propose a novel multi-modal method leveraging audio-lyrics contrastive loss and two symmetric cross-modal attention, to align and fuse the features from audio and lyrics.

In order to improve the performance of multi-label MGC, in addition to using the multi-modality nature of music, we can also start from the nature of the multi-label problem.
To address the multi-label classification problem, the easiest way is to treat each genre individually.
However, this approach is inherently limited because it ignores the potential genre correlations which can be approximated as co-occurrence and mutual exclusion. As this correlation is essentially a topological structure, we design a genre correlations extraction module to capture and model it. 

This paper has the following main contributions:
1) we propose a novel multi-modal approach leveraging audio-lyrics contrastive loss and two symmetric cross-modal attention, to align and fuse the features from audio and lyrics;
2) based on the nature of the multi-label classification problem, we design a genre correlations extraction module which can capture and model the potential genre correlations. 
3) our proposed method significantly surpasses the previous methods and achieves state-of-the-art performance on Music4All dataset.

\begin{figure*}[ht]
	\centering
	\includegraphics[width=0.90\linewidth]{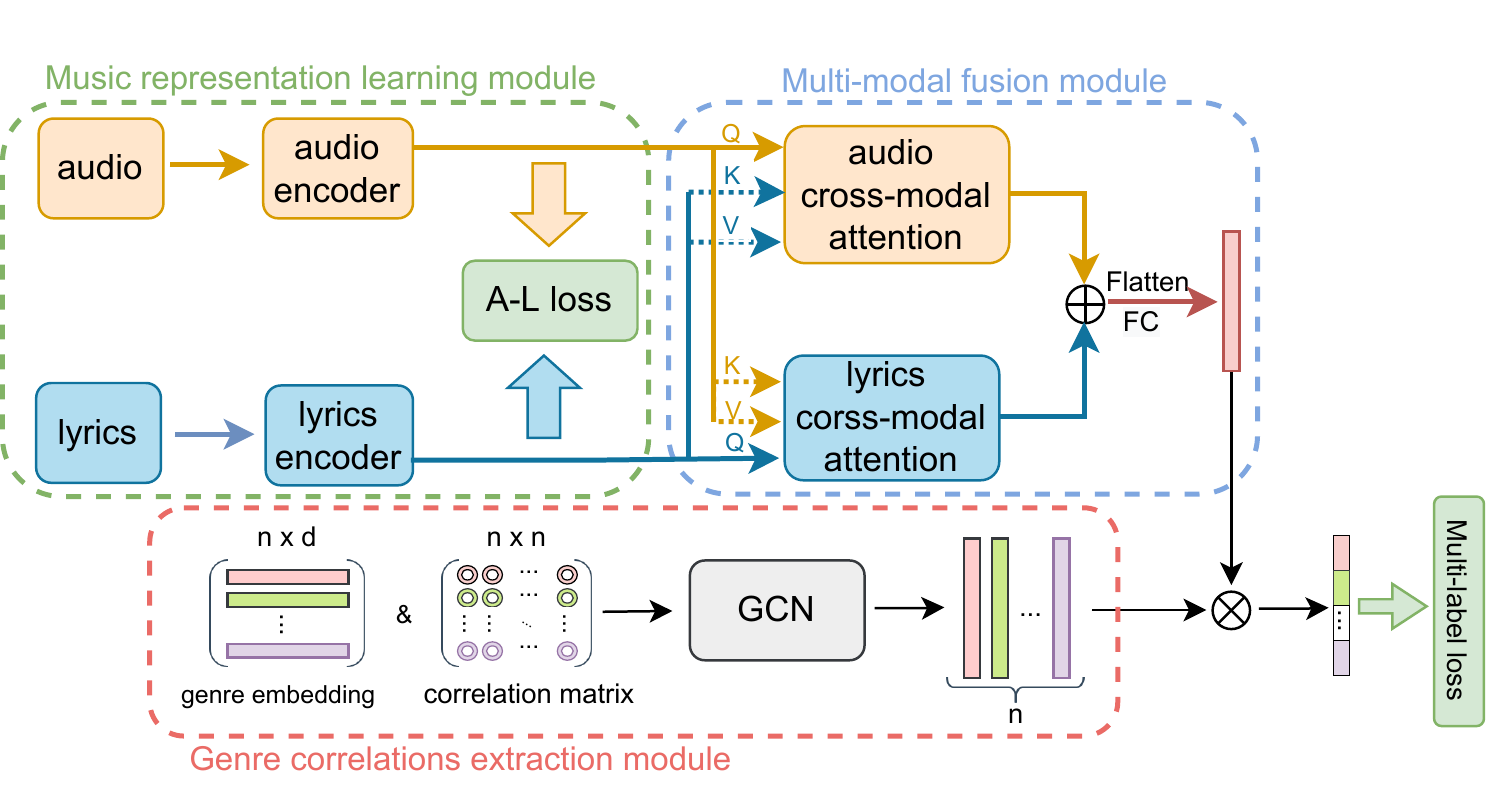}	
	\caption{Architectures of our proposed model, which consists of three modules: music representation learning module, multi-modal fusion module, and genre correlations extraction module.
 }
	\label{figure1}
\end{figure*}

\section{Method}
\label{sec:format}

% The music representation learning module extracts features from audio and lyrics modalities respectively. Furthermore, an audio-lyrics contrastive loss (A-L loss) is applied to align features. The multi-modal fusion module consists of two symmetric cross-modal attention and efficiently fuses features from different modalities. The genre correlations extraction module uses a graph convolution network (GCN) to model potential correlations between genres.

As illustrated in Figure \ref{figure1}, our model consists of three main modules: music representation learning module, multi-modal fusion module, and genre correlations extraction module. 
The music representation learning module extracts features from audio and lyrics modalities respectively. Furthermore, an audio-lyrics contrastive loss (A-L loss) is applied to align features obtained from different modalities before sending them into the multi-modal fusion module.
The multi-modal fusion module is composed of two symmetrical mechanisms and efficiently fuses different features.
The genre correlations extraction module starts from the nature of the multi-label problem and uses a graph convolution network (GCN) to model potential correlations between genres. Finally, a classifier takes the dot-product of the fused feature representation and the embeddings of genre categories as input to predict genres.
% Specifically, 
% Given music metadata (audio and lyrics), our proposed approach first extracts feature representations respectively by music representation learning module and then applies two symmetric cross-modal attention to fuse them. Besides, a graph convolution network module starts from the nature of the multi-label problem to extract potential genre correlations. Finally, a classifier takes the dot-product of the fused feature representation and the embeddings of genre categories as input to predict genres.

\subsection{Music Representation Learning Module}

The music representation learning module consists of an audio encoder and a lyrics encoder, which takes audio and lyrics as input and outputs corresponding features respectively. According to the pre-processing procedure adopted in \cite{audio_encoder2}, the audio is cropped to 30 seconds and preprocessed to a Mel-spectrogram with 22050HZ sampling rate, 512 hop-size, and 128 Mel-filters. Compared with waveform audio, Mel-spectrogram has two advantages: 1) it is designed based on the theory of human aural perception
and instrument frequency range; 2) it can significantly reduce the amount of data that the network needs to process. 

Considering the powerful feature extraction capabilities of convolutional neural networks, 
we refer to \cite{zhao} to develop the audio encoder that consists of five convolutional blocks, whose input is the preprocessed Mel-spectrogram and output is the abstract audio features. 
% Each convolution block is composed of convolution layer, batch normalization layer and max-pooling layer in series.
% The ReLU function is used as a non-linearity after batch 
% normalization layer.
% audio encoder consists of five convolutional blocks 
% which contain a 2D convolutional layer, a batch normalization layer, and a 2D max-pooling layer.
% The kernel size of convolutional layer is set to 3*3, and the stride is set to 1. 
% The max-pooling layer shares a window size of 2x2 to reduce the feature map size. 
% The five convolutional layers have 32, 64, 128, 256, and 512 filters respectively. The kernel size of filters is set to 3*3, and the stride is set to 1. All max-pooling layers share a window size of 2x2 to reduce the feature map size.

As we all know, BERT model \cite{bert} has a strong ability to extract text information, which has been confirmed in many NLP tasks. Therefore, BERT model is used as the lyrics encoder and its structure is not changed or improved. Considering that BERT model requires a large amount of text corpus for training, we directly use the pre-trained BERT model and freeze its parameters.

\subsection{Multi-modal Fusion Module}

It is crucial to effectively fuse features from different modalities in multi-modal approaches. 
Since it is difficult for methods based on concatenation or element addition to make the information interaction between different modalities, we design a novel multi-modal fusion module which consists of two symmetric cross-modal attention to effectively fuse multi-modal features. The cross-modal attention is an extended application of self-attention \cite{self_attention} on multi-modal tasks, which emphasizes important parts of features by computing similarity between different features. The calculation formula of cross-modal attention is:
\begin{equation}
\label{E1}
Attention(Q_\alpha, K_\beta, V_\beta) = softmax(\frac{Q_{\alpha}K^{T}_{\beta}}{\sqrt{d}})V_{\beta}
\end{equation}
where $\alpha$ and $\beta$ denote different modalities. 
% \textit{Q}, \textit{K}, \textit{V} represent \textit{Query}, \textit{Key} and \textit{Value} respectively, \textit{d} is the dimension size, $\alpha$ and $\beta$ denote different modalities. 
%In practical scenarios, to enhance the learning ability, multi-head attention is applied, and it can be defined as :
% \begin{equation}
% \label{E2}
% head_i = Attention(QW_{i}^{Q},KW_{i}^{K}, VW_{i}^{V})
% \end{equation}
% \begin{equation}
% \label{E3}
% MHA(Q,K,V) = Concat(head_{1}, .... , head_{n})W^{O}
% \end{equation}
% where \textit{$W^{Q}$}, \textit{$W^{K}$}, \textit{$W^{V}$} are the learnable weight matrix of \textit{Query}, \textit{Key} and \textit{Value} respectively, and \textit{$W^{O}$} is another learnable weight matrix which can adaptively combine single-head attention. 
Considering that both audio and lyrics contribute significantly to genre classification, our proposed two symmetric cross-modal attention consists of audio cross-modal attention and lyrics cross-modal attention. In audio cross-modal attention, \textit{Query} is obtained from audio feature, and both \textit{Key} and \textit{Value} are obtained from lyrics feature. It is the opposite in lyrics cross-modal attention.

\subsection{Genre Correlations Extraction Module}

How to effectively capture and model the correlations between genres to improve the classification performance is especially important for multi-label MGC task. Considering that this correlation is essentially a topological structure, we design a graph convolution network \cite{gcn} which is well suited to handle discrete topological data to model the genre correlations. The graph convolution network takes a set of nodes and the corresponding adjacency matrix as inputs, and outputs the updated nodes.
% $ A \in \mathbb{R}^{n * n}$ as inputs (where \textit{n} is the number of nodes and \textit{d} denotes the dimensionality of node features), and outputs the updated nodes $ \hat{H} \in \mathbb{R}^{n * d}$.
% which can be defined as : 
% \begin{equation}
% \label{E4}
% \begin{align*}
% \hat{H} & = GCN(H, A) \\
%         & = \sigma(\tilde{D}^{-\frac{1}{2}}A\tilde{D}^{-\frac{1}{2}}HW)
% \end{align*}
% \end{equation}
% where $\tilde{D}$ is the degree matrix of \textit{A}, \textit{W} is a learnable weight matrix, and $\sigma$ is the relu activation function. 
A pre-trained BERT \cite{bert} model is used to encode genre names as semantic embedding and treat them as genre node features.
The correlation matrix is obtained by calculating the co-occurrence conditional probability, and its calculation formula is:
\begin{equation}
\label{E5}
A_{ij}^{(1)} = P(L_{j} | L_{i}) = \frac{N(L_{i}L_{j})}{N(L_{j})}
\end{equation}
where $N(L_{j})$ is the number of occurrences of genre \textit{j}, and $N(L_{i}L_{j})$ denotes the concurring times of genre \textit{i} and genre \textit{j}. 
However, this statistical conditional probability is determined by the distribution of samples in the training set, which can be influenced significantly by noise and class imbalance. 
Genre names have inherent textual information because they are summed up based on musical and cultural understandings. Therefore, we utilise the similarity matrix to alleviate the shortcomings of the co-occurrence conditional probability matrix. The similarity matrix is obtained by computing the cosine similarity between node features, which represents the proximity of genre names in the semantic space. The calculation formula of similarity matrix is as follows:
\begin{equation}
\label{E6}
A_{ij}^{(2)} = cos(F_{i}, F_{j}) = \frac{F_{i} * F_{j}}{{||F_{i}||}_{2} \cdot {||F_{j}||}_{2}}
\end{equation}
where $F_{i}$ and $F_{j}$ denote the feature of node \textit{i} and node \textit{j}, and  $cos(\cdot)$ is the cosine similarity function. The correlation matrix is then defined as $A = \frac{1}{2}(A^{(1)} + A^{(2)})$.

% \begin{equation}
% \label{E7}
% A = \frac{1}{2}(A^{(1)} + A^{(2)})
% \end{equation}

\subsection{Contrastive Audio-lyrics Alignment}
In MGC task, existing multi-modal methods directly fuse features obtained from different modalities without considering the alignment between these modalities. However, These methods lead to less satisfactory fusion because they ignore that features from different modalities are inherently heterogeneous and misaligned.
To address this issue, we propose an audio-lyrics contrastive loss to align features obtained from different modalities. Specifically, given the embedding of audio and lyrics, we optimize a similarity function $s(\cdot)$ between audio \textit{A} and lyrics \textit{L}:
\begin{equation}
\label{E8}
s(A,L) = g_{a}(E_{a}) \cdot g_{l}(E_{l})
\end{equation}
where $g_{a}(\cdot)$ and $g_{l}(\cdot)$ are linear projection layers which transform the embedding $E_{a}$ and $E_{l}$ to a low-dimensional space.
Following \cite{contrastive_loss1, contrastive_loss2, contrastive_loss3}, for each input audio-lyrics pair $\left\langle A_{i}, L_{i} \right\rangle$, the audio-lyrics contrastive loss is composed of two symmetric terms, audio-to-lyrics term:
\begin{equation}
\label{E8}
\mathcal{L}_{A \to L} = -\log\frac{\exp(s(A_{i},L_{i})/\tau)}{\sum_{j=1}^B\exp(s(A_{i},L_{j})/\tau)}
\end{equation}
and lyrics-to-audio term:
\begin{equation}
\label{E9}
\mathcal{L}_{L \to A} = -\log\frac{\exp(s(L_{i},A_{i})/\tau)}{\sum_{j=1}^B\exp(s(L_{i},A_{j})/\tau)}
\end{equation}
where B is the batch size and $\tau$ is a adaptive parameter. The audio-lyrics contrastive loss is then defined as $\mathcal{L}_{A-L} = \frac{1}{2}(\mathcal{L}_{A \to L} + \mathcal{L}_{L \to A})$.

To enable the whole model to be trained end-to-end, we also adopted the commonly used BCE loss as the multi-label loss. The final training loss is the weighted sum of $\mathcal{L}_{A-L}$ and $\mathcal{L}_{BCE}$, which can be defined as:
\begin{equation}
\label{E10}
\mathcal{L} = \lambda\mathcal{L}_{A-L} + (1-\lambda)\mathcal{L}_{BCE}
\end{equation}
where $\lambda$ is a hyperparameter that is used to balance $\mathcal{L}_{A-L}$ and $\mathcal{L}_{BCE}$.

\section{experiments}
\subsection{Experimental Setup}
We use Music4All dataset \cite{music4all} to train and test our model. Considering that Music4All dataset is noisy and contains multiple languages in music lyrics, we filtered out music tracks with missing information or non-English lyrics in our experiment. 43650 samples with 87 genre categories are finally selected and are split into 3 subsets with approximately 70\%, 10\%, and 20\% samples respectively. In the experiment, we set the learning rate to 0.0001
and decays by half every 50 epochs, and set the batch size to 16.
% In addition, the Adam optimizer is adopted to train our model.
% we adopted the Adam optimizer with a weight decay of 0.02 and set the batch size to 16. The initial learning rate is set to 0.0001 and decays by half every 50 epochs. 
Following \cite{audio_encoder2, music4all}, accuracy and F-measure are adopted to evaluate our proposed model.
% Although accuracy is easily affected by class imbalance, it can intuitively reflect the classification performance of the model. Besides, F-measure is adopted, which comprehensively considers the precision rate and the recall rate to further explore the model performance.
% The formula for calculating the F-measure is as follows:
% \begin{equation}
% \label{E11}
% F = \frac{2 * P * R}{P+R}
% \end{equation}
% where \textit{P} and \textit{R} denote the precision rate and the recall rate respectively.

\subsection{Results and Analysis}
In order to compare fairly with other multi-label MGC methods, we reproduced Pandeya's model \cite{audio_encoder2} and Santana's model \cite{music4all} on the filtered dataset with the same experimental settings. The comparison results with other multi-modal methods are shown in Table \ref{table1}. Our model achieves 0.456 in accuracy and 0.534 in F-measure, which significantly outperforms previous models by a large margin. Santana's model achieves the worst results because it only used audio information, which indicates that multi-modal information is crucial for improving the classification performance of multi-label MGC methods. Pandeya's model uses both audio information and lyrics information as input and shows the state-of-art performance in multi-label MGC task. However, this method does not achieve substantial performance gains (only from 0.354 to 0.387) in accuracy. This is mainly because it simply concatenates the features from different modalities, and does not consider the differences between modalities.

% As there are few pieces of research on multi-label MGC, we cannot conduct comparative experiments with more methods. From the above results, our model surpasses two previous models and achieves state-of-the-art result on Music4All dataset.

\begin{table}
    \centering
    \caption{Comparison with other multi-modal methods on Music4All dataset.}
    \label{table1}
    \begin{tabular}{c|cc}
        \hline
        Method & Accuracy & F-measure\\
        \hline
        Santana’s model \cite{music4all}  & 0.354 & 0.419 \\
        Pandeya's model \cite{audio_encoder2}  & 0.387 & 0.442 \\
        Our proposed model  & 0.456 & 0.534 \\
        \hline
    \end{tabular}
\end{table}

\begin{table}
    \centering
    \caption{Ablation studies on three components. A-L loss, SCMA, and GCEM denote audio-lyrics contrastive loss, two symmetric cross-modal attention, and genre correlations extraction module respectively.}
    \label{table2}
    \begin{tabular}{ccc|cc}
        \hline
        A-L loss & SCMA & GCEM  & Accuracy & F-measure\\
        \hline
         & & & 0.372 & 0.438 \\
        \checkmark & & & 0.396 & 0.475 \\
         & \checkmark & & 0.408 & 0.491 \\
         & & \checkmark & 0.403 & 0.487 \\
        \checkmark & \checkmark & & 0.425 & 0.513 \\
        \checkmark & & \checkmark & 0.419 & 0.508 \\
         & \checkmark & \checkmark & 0.437 & 0.515 \\
        \checkmark & \checkmark & \checkmark & 0.456 & 0.534 \\
        \hline
    \end{tabular}
\end{table}

\begin{figure}[htbp]
	
	\begin{minipage}{0.47\linewidth}
		\vspace{3pt}
        %这个图片路径替换成你的图片路径即可使用
		\centerline{\includegraphics[width=\textwidth]{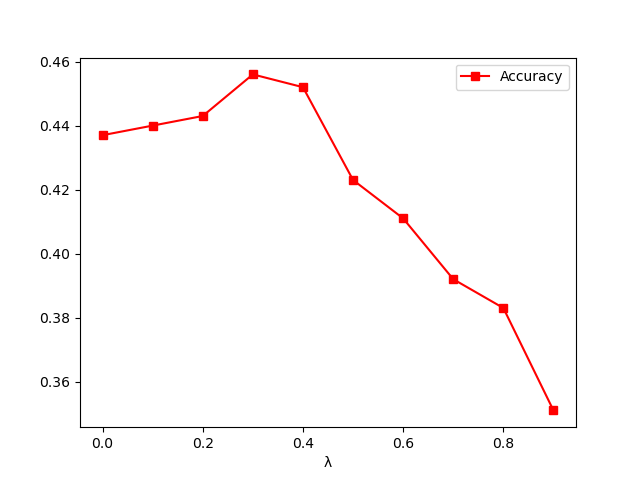}}
          % 加入对这列的图片说明
		\centerline{(a) accuracy}
	\end{minipage}
	\begin{minipage}{0.47\linewidth}
		\vspace{3pt}
		\centerline{\includegraphics[width=\textwidth]{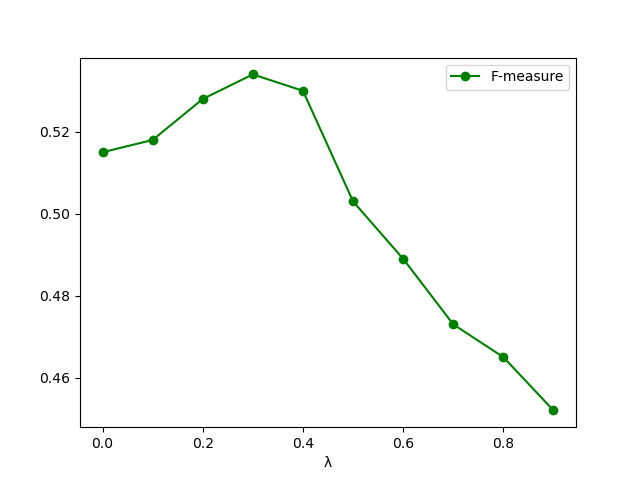}}
	 
		\centerline{(b) F-measure}
	\end{minipage}
	\caption{The accuracy and F-measure of our proposed model when using different $\lambda$. The horizontal and vertical coordinates represent the value of $\lambda$ and the performance of the model, respectively.}
	\label{fig2}
\end{figure}

\subsection{Ablation Study}
\emph{1) three key components:} 
We conducted a set of ablation experiments to evaluate three key components: audio-lyrics contrastive loss (A-L loss), symmetric cross-modal attention (SCMA), and genre correlations extraction module (GCEM).

Table \ref{table2} reports the ablation results. We find that the baseline model without any components shows the worst performance, only achieving 0.372 in accuracy and 0.438 in F-measure. In contrast, no matter which component is added, the performance of the model increases accordingly, indicating that all three components contribute to the model’s performance. Interestingly, we observe that adding A-L loss or SCMA achieves 0.396 or 0.408 in accuracy respectively, which both surpass the baseline model. This implies that both feature alignment and effective fusion strategies help to eliminate disparity. In addition, it is notable that additional GCEM brings considerable improvement in the relevant ablation experiments, justifying the exploration of the correlation between genres. In all ablation experiments, we find that the most complete model achieves the best results, indicating that the three components can be combined flexibly without conflict.

\emph{2) different $\lambda$ in training loss:} Figure \ref{fig2} shows the performance of our proposed model when using different $\lambda$ in Eq. \ref{E10}. We find that with the change of $\lambda$, both accuracy and F-measure increased first and then decreased, and achieved the maximum value when $\lambda$=0.3. This demonstrates that proper A-L loss can help improve the performance of classification. It is notable that the model fails to converge when $\lambda$=1.0, which shows that the multi-label loss is crucial to the convergence of the model.

\section{conclusion}
In this paper, we improve multi-label music genre classification from multi-modal properties of music and genre correlations perspective and have presented a novel multi-modal method leveraging audio-lyrics contrastive loss and two symmetric cross-modal attention, to align and fuse features from audio and lyrics. Furthermore, a graph convolution network is proposed to capture and model potential genre correlations. 
Extensive experiments demonstrate that our proposed method significantly surpasses other multi-label MGC methods and achieves state-of-the-art result on Music4All dataset.
% A series of ablation experiments have proved that all three components contribute to improving classification performance and can be combined flexibly without conflict.

\section{Acknowledgement}
Supported by the Key Research and Development Program of Guangdong Province (grant No. 2021B0101400003) and Corresponding author is Jianzong Wang (jzwang@188.com).
% This paper is supported by the Key Research and Development Program of Guangdong Province under grant No.2021B-
% 0101400003. Corresponding author is Jianzong Wang from Ping An Technology (Shenzhen) Co., Ltd (jzwang@188.com).

% \section{REFERENCES}
% \label{sec:refs}

\bibliographystyle{IEEEbib}
\bibliography{paper}

\begin{thebibliography}{10}

\bibitem{early_work1}
Michael~I. Mandel and Daniel P.~W. Ellis,
\newblock ``Song-level features and support vector machines for music
  classification,''
\newblock in {\em International Society for Music Information Retrieval
  Conference}, 2005.

\bibitem{early_work3}
James Bergstra, Norman Casagrande, Dumitru Erhan, Douglas Eck, and Bal{\'a}zs
  K{\'e}gl,
\newblock ``Aggregate features and adaboost for music classification,''
\newblock {\em Machine learning}, vol. 65, no. 2, pp. 473--484, 2006.

\bibitem{lyrics}
Chia-Hung Yeh, Wen-Yu Tseng, Chia-Yen Chen, Yu-Dun Lin, Yi-Ren Tsai, Hsuan-I
  Bi, Yu-Ching Lin, and Ho-Yi Lin,
\newblock ``Popular music representation: chorus detection \& emotion
  recognition,''
\newblock {\em Multimedia tools and applications}, vol. 73, no. 3, pp.
  2103--2128, 2014.

\bibitem{cnn}
Nikki Pelchat and Craig~M Gelowitz,
\newblock ``Neural network music genre classification,''
\newblock {\em Canadian Journal of Electrical and Computer Engineering}, vol.
  43, no. 3, pp. 170--173, 2020.

\bibitem{rnn}
Zain Nasrullah and Yue Zhao,
\newblock ``Music artist classification with convolutional recurrent neural
  networks,''
\newblock in {\em 2019 International Joint Conference on Neural Networks
  (IJCNN)}. IEEE, 2019, pp. 1--8.

\bibitem{multi-modal1}
Sergio Oramas, Francesco Barbieri, Oriol Nieto~Caballero, and Xavier Serra,
\newblock ``Multimodal deep learning for music genre classification,''
\newblock {\em Transactions of the International Society for Music Information
  Retrieval. 2018; 1 (1): 4-21.}, 2018.

\bibitem{multi-modal2}
Rafael~B Mangolin, Rodolfo~M Pereira, Alceu~S Britto, Carlos~N Silla,
  Val{\'e}ria~D Feltrim, Diego Bertolini, and Yandre~MG Costa,
\newblock ``A multimodal approach for multi-label movie genre classification,''
\newblock {\em Multimedia Tools and Applications}, pp. 1--26, 2020.

\bibitem{multi-modal3}
Wanglei Shi and Shuang Feng,
\newblock ``Research on music emotion classification based on lyrics and
  audio,''
\newblock in {\em 2018 IEEE 3rd Advanced Information Technology, Electronic and
  Automation Control Conference (IAEAC)}. IEEE, 2018, pp. 1154--1159.

\bibitem{zhao}
Jiahao Zhao, Ganghui Ru, Yi~Yu, Yulun Wu, Dichucheng Li, and Wei Li,
\newblock ``Multimodal music emotion recognition with hierarchical cross-modal
  attention network,''
\newblock in {\em 2022 IEEE International Conference on Multimedia and Expo
  (ICME)}. IEEE, 2022, pp. 1--6.

\bibitem{six_feature}
Igor Vatolkin and Cory McKay,
\newblock ``Multi-objective investigation of six feature source types for
  multi-modal music classification,''
\newblock {\em Transactions of the International Society for Music Information
  Retrieval}, vol. 5, no. 1, 2022.

\bibitem{audio_encoder2}
Yagya~Raj Pandeya, Jie You, Bhuwan Bhattarai, and Joonwhoan Lee,
\newblock ``Multi-modal, multi-task and multi-label for music genre
  classification and emotion regression,''
\newblock in {\em 2021 International Conference on Information and
  Communication Technology Convergence (ICTC)}. IEEE, 2021, pp. 1042--1045.

\bibitem{bert}
Jacob Devlin, Ming-Wei Chang, Kenton Lee, and Kristina Toutanova,
\newblock ``Bert: Pre-training of deep bidirectional transformers for language
  understanding,''
\newblock {\em arXiv preprint arXiv:1810.04805}, 2018.

\bibitem{self_attention}
Ashish Vaswani, Noam Shazeer, Niki Parmar, Jakob Uszkoreit, Llion Jones,
  Aidan~N Gomez, {\L}ukasz Kaiser, and Illia Polosukhin,
\newblock ``Attention is all you need,''
\newblock {\em Advances in neural information processing systems}, vol. 30,
  2017.

\bibitem{gcn}
Thomas Kipf and Max Welling,
\newblock ``Semi-supervised classification with graph convolutional networks,''
\newblock {\em ArXiv}, vol. abs/1609.02907, 2016.

\bibitem{contrastive_loss1}
Pin Jiang and Yahong Han,
\newblock ``Reasoning with heterogeneous graph alignment for video question
  answering,''
\newblock in {\em Proceedings of the AAAI Conference on Artificial
  Intelligence}, 2020, vol.~34, pp. 11109--11116.

\bibitem{contrastive_loss2}
Alec Radford, Jong~Wook Kim, Chris Hallacy, Aditya Ramesh, Gabriel Goh,
  Sandhini Agarwal, Girish Sastry, Amanda Askell, Pamela Mishkin, Jack Clark,
  et~al.,
\newblock ``Learning transferable visual models from natural language
  supervision,''
\newblock in {\em International Conference on Machine Learning}. PMLR, 2021,
  pp. 8748--8763.

\bibitem{contrastive_loss3}
Dongxu Li, Junnan Li, Hongdong Li, Juan~Carlos Niebles, and Steven~CH Hoi,
\newblock ``Align and prompt: Video-and-language pre-training with entity
  prompts,''
\newblock in {\em Proceedings of the IEEE/CVF Conference on Computer Vision and
  Pattern Recognition}, 2022, pp. 4953--4963.

\bibitem{music4all}
Igor Andr{\'e}~Pegoraro Santana, Fabio Pinhelli, Juliano Donini, Leonardo
  Catharin, Rafael~Biazus Mangolin, Val{\'e}ria~Delisandra Feltrim,
  Marcos~Aur{\'e}lio Domingues, et~al.,
\newblock ``Music4all: A new music database and its applications,''
\newblock in {\em 2020 International Conference on Systems, Signals and Image
  Processing (IWSSIP)}. IEEE, 2020, pp. 399--404.

\end{thebibliography}

\end{document}